\documentclass[prd,aps,twocolumn,showpacs]{revtex4}
\usepackage{latexsym}
\usepackage{amssymb}
\begin{document}
\newcommand{\beq}{\begin{equation}}
\newcommand{\eeq}{\end{equation}}
\newcommand{\beqa}{\begin{eqnarray}}
\newcommand{\eeqa}{\end{eqnarray}}
\newcommand{\Prd}{Phys. Rev D}
\newcommand{\Prl}{Phys. Rev. Lett.}
\newcommand{\Plb}{Phys. Lett. B}
\newcommand{\Cqg}{Class. Quantum Grav.}
\newcommand{\Np}{Nuc. Phys.}
\newcommand{\Ap}{Ann. Phys.}
\newcommand{\Ptp}{Prog. Theor. Phys.}
\newcommand{\half}{\frac 1 2\;}
\newcommand{\dg}{\digamma}

\title{Causality and charged spin-2 fields in an electromagnetic background}
\author{M. Novello, S. E. Perez Bergliaffa, and R. P. Neves}
\affiliation{ Centro Brasileiro de Pesquisas F\'{\i}sicas\\Rua
Dr.\ Xavier Sigaud 150, Urca 22290-180\\ Rio de Janeiro,
Brazil}
\date{\today}

\begin{abstract}
We show that, contrary to common belief,
the propagation of a spin-2 field in an electromagnetic background is
{\em causal}. The proof will be given in the Fierz formalism which, as we shall see,
is free of the ambiguity present in the more usual Einstein representation.

\end{abstract}


\maketitle

 \renewcommand{\thefootnote}{\arabic{footnote}}

\section{Introduction}


The interaction of fields with spin greater than 1 (particularly $s=2$ fields)
with a fixed
gravitational or electromagnetic background has attracted
a lot of attention in the last three decades.
There are at least two reasons for this interest.
First, on the theoretical side,
interacting particles with $s>1$ present
features that are absent in Electromagnetism.
In this regard, two
items are specially important: the consistency of the
equations of motion (EOM), and the causality of the propagation.
A consistent set of EOM (and of the constraints derived from them)
has been obtained quite a long time ago for free fields (see for instance
\cite{sh}), but the consistency of the system is usually broken when interactions
are introduced. Interactions can also excite new
degrees of freedom, absent in the free-field case. This may lead
to violation of causality. In fact, it has been stated that causality can be
violated even when the higher spin fields have the correct number of degrees
of freedom, and that in order to ensure consistency, some restrictions
must be made on the kind of interaction \cite{r1}.

Second, on a more phenomenological vein,
particles
with spin 2 are known to exist as resonances, and it is desirable to have a theory
to describe their interaction with background fields. Yet another reason is furnished by
string theory, in which a tower of massive states of Kaluza-Klein type
with various spins
interact with each other.
A particular instance
of this scenario is furnished by
the so-called bigravity models \cite{damour1, damour2}, from which
a theory with a massless and a very light graviton can be obtained.
It would be very interesting then to have a consistent
theory of fields with $s>1$ (and specifically of $s=2$ fields)
interacting with given backgrounds.

A lot of work has been devoted to the case of a gravitational
background. The properties of an $s=2$ field in this kind of
background were studied for instance in
\cite{aragones71,r1,r2,d2}. Several interesting new results in
this area were obtained in \cite{nn}. We shall analyze here
instead a spin 2 field in an electromagnetic (EM) background
\cite{cou}. It is common lore in this situation that massive spin
2 particles propagate {\em acausally}. This result was obtained by
Kobayashi and Shamaly in the late 70's \cite{ks}, and a more
recent demonstration has been given by Deser and Waldron
\cite{ds}, both proofs being based on the method of
characteristics. The main result we shall present here is that,
contrary to the aforementioned claims, a more careful application
of the method of characteristics reveals that the propagation of
$s=2$ fields in an EM background is actually {\em causal}.

Let us remind the reader that a spin-2 field can be described in
two equivalent ways, which we shall call the Einstein
representation (ER) and the Fierz representation (FR). The former
is a second order representation that uses a symmetric
second-order tensor $\varphi_{\mu\nu}$ to represent the field. In
the FR \cite{fp}, this role is played by a third order tensor
$F_{\alpha\mu\nu},$ which is antisymmetric in the first pair of
indices, and obeys the cyclic identity and a further condition
(which will be given in Sect.\ref{fabc}) in order to represent a
single spin-2 field. The FR is first order. In flat spacetime and
in the absence of interactions both representations are
equivalent. Nevertheless, in the case an EM background (or in
curved spacetime) this is no longer true. As we shall see, in the
ER there is an ambiguity which originates in the ordering of the
non-commuting covariant derivatives. We shall show that the use of
the FR yields instead a non-ambiguous description with the minimal
coupling procedure.

Our proof of the causal propagation will be given
in the FR.
We shall begin by giving in Sect.\ref{fabc}
a review of the Fierz variables to describe a spin-2
field in Minkowski spacetime (some properties of these variables are
discussed in Appendix 1). In Sect.\ref{int} it will be shown how the minimal coupling
in the FR avoids the ambiguity present in the ER. The causality of the propagation
will be discussed in Sect.\ref{caus}. We close with a discussion of the results.

\section{Spin-2 field description in the Fierz representation}
\label{fabc}

In this section we present a short review of the FR in
a Minkowskian background and in the absence of interactions
\footnote{Throughout this article we will use
the signature $(+---)$, and the notation $A_{(\alpha}B_{\beta )} = A_\alpha B_\beta
+ A_\beta B_\alpha$, $A_{[\alpha}B_{\beta ]} = A_\alpha B_\beta
- A_\beta B_\alpha$.}. Let us start by
defining a three-index tensor $F_{\alpha\beta\mu}$ which is
anti-symmetric in the first pair of indices and obeys the cyclic
identity:
\begin{equation}
F_{\alpha\mu\nu} + F_{\mu\alpha\nu} = 0,
\label{01}
\end{equation}
\begin{equation}
F_{\alpha\mu\nu} + F_{\mu\nu\alpha} + F_{\nu\alpha\mu} = 0.
\label{02}
\end{equation}
The former expression implies that the dual of $F_{\alpha\mu\nu}$ is
trace-free:
\begin{equation}
\stackrel{*}{F}{}^{\alpha\mu}{}_{\mu} = 0 ,
\label{02bis}
\end{equation}
where the asterisk represents the dual operator, defined in terms
of $\eta_{\alpha\beta\mu\nu}$ by
\[
\stackrel{*}{F}{}^{\alpha\mu}{}_{\lambda} \equiv \frac{1}{2} \,
\eta^{\alpha\mu}{}_{\nu\sigma}\,F^{\nu\sigma}{}_{\lambda}.
\]
The tensor $F_{\alpha\mu\nu}$ has 20 independent components.
The necessary and sufficient condition for $F_{\alpha\mu\nu}$ to
represent an unique spin-2 field (described by 10 components)
is \footnote{Note that this
condition is analogous to that necessary for
the existence of a potential
$A_{\mu}$ for the EM field, given by
$\stackrel{*}{A}{}^{\alpha\mu}{}_{,\alpha} = 0.$}
\begin{equation}
\stackrel{*}{F}{}^{\alpha (\mu\nu)}{}_{,\alpha} = 0,
\label{03}
\end{equation}
which can be rewritten as
\begin{eqnarray}
&& {{F_{\alpha\beta}}^{\lambda}}{}_{,\mu} + {{F_{\beta\mu}}^{\lambda}}%
{}_{,\alpha} + {{F_{\mu\alpha}}^{\lambda}}{}_{,\beta} -\frac{1}{2}
\delta^{\lambda}_{\alpha} (F_{\mu ,\beta} - F_{\beta ,\mu}) + \nonumber \\
&& - \frac{1}{2} \delta^{\lambda}_{\mu} (F_{\beta ,\alpha} - F_{\alpha
,\beta}) - \frac{1}{2} \delta^{\lambda}_{\beta} (F_{\alpha ,\mu} - F_{\mu
,\alpha}) = 0.
\end{eqnarray}
A direct consequence of the above
equation is the identity:
\begin{equation}
F^{\alpha\beta\mu}{}_{\, ,\mu} = 0 \ .  \label{z1}
\end{equation}
We will call a tensor that satisfies the conditions given in the Eqns.(\ref{01}),
(\ref{02}) and (\ref{03}) a Fierz tensor.

If $F_{\alpha\mu\nu}$ is a Fierz tensor, it represents a
unique spin-2 field. Condition (\ref{03}) yields a
connection between the ER and the FR: it implies that
there exists a symmetric second-order tensor $\varphi_{\mu\nu}$
such that
\begin{eqnarray}
2\,F_{\alpha\mu\nu} &=& \varphi_{\nu [\alpha,\mu ]} +
\left( \varphi_{,\alpha} - \varphi_{\alpha}{}^{\lambda}{}_{,\lambda}
\right)\, \eta_{\mu\nu}\nonumber \\
 &-& \left(\varphi_{,\mu} -
\varphi_{\mu}{}^{\lambda}{}_{,\lambda} \right)\, \eta_{\alpha\nu} . \label{04}
\end{eqnarray}
where $\eta_{\mu\nu}$ is the flat spacetime metric tensor, and
the factor $2$ in the l.h.s. is introduced for convenience.

Taking the trace of equation (\ref{04}) we find that
$$
F_{\alpha} = \varphi_{,\alpha} - \varphi_{\alpha}{}^{\lambda}{}_{,\lambda},
$$
where $F_{\alpha}\equiv F_{\alpha\mu\nu}\eta^{\mu\nu}$. Thus we can write
\begin{equation}
2 F_{\alpha\mu\nu} = \varphi_{\nu [\alpha,\mu ]}  +
F_{[\alpha } \,\eta_{\mu ]\nu}.  \label{06}
\end{equation}

The following identity can proved using
the properties of the Fierz tensor:
\begin{equation}
F^{\alpha }{}_{(\mu \nu ),\alpha }\equiv -\,G^{(L)}{}_{\mu \nu } ,
\label{07}
\end{equation}
where $G^{(L)}{}_{\mu \nu }$ is the linearized Einstein tensor, defined in
terms of the symmetric tensor $\varphi _{\mu \nu }$ by
\begin{equation}
G^{(L)}{}_{\mu \nu }\equiv \Box \,\varphi _{\mu \nu }-\varphi ^{\epsilon
}{}_{(\mu ,\nu )\,,\epsilon }+\varphi _{,\mu \nu }-\eta _{\mu \nu }\,\left(
\Box \varphi -\varphi ^{\alpha \beta }{}_{,\alpha \beta }\right) . \label{08}
\end{equation}

The divergence of $F^{\alpha }{}_{(\mu \nu ),\alpha }$ yields the identity:

\begin{equation}
F^{\alpha (\mu \nu )}{}_{,\alpha \mu }\equiv 0.  \label{07bis}
\end{equation}
Indeed,
\begin{equation}
F^{\alpha \mu \nu }{}_{,\alpha \mu }+F^{\alpha \nu \mu }{}_{,\mu \alpha }=0.
\label{070}
\end{equation}
The first term vanishes identically due to the symmetry properties
of the field and the second term vanishes due to equation
(\ref{z1}). Using Eqn.(\ref{07})
the identity which states that the
linearized Einstein tensor $G^{(L)}{}_{\mu \nu }$ is
divergence-free is recovered.

We shall build now dynamical equations for the free Fierz tensor in flat spacetime.
Our considerations will be restricted here to linear dynamics \cite{mio}.
The most general
theory can be constructed from a combination of the three invariants involving
the field. These are
represented by $A$, $B$ and $W$:
$$
A \equiv F_{\alpha \mu \nu }\hspace{0.5mm}F^{\alpha \mu \nu } ,
\;\;\;\;\;\;\;\;B \equiv F_{\mu }\hspace{0.5mm}F^{\mu },
$$
$$
W \equiv F_{\alpha \beta \lambda }\stackrel{\ast }{F}{}^{\alpha \beta
\lambda }=\frac{1}{2}\,F_{\alpha \beta \lambda }\hspace{0.5mm}F^{\mu \nu
\lambda }\,\eta ^{\alpha \beta }{}_{\mu \nu } .
$$
$W$ is a topological
invariant in the linear regime, so we shall use in what follows only
the invariants $A$ and $B$.

The EOM  for the massless spin-2 field in the ER is given by

\begin{equation}
G^{(L)}{}_{\mu \nu }=0.  \label{014bis}
\end{equation}
As we have seen above, in terms of the field $F^{\lambda \mu \nu}$
this equation can be written as

\begin{equation}
F^{\lambda (\mu \nu )}{}_{,\lambda }=0.  \label{014}
\end{equation}
The corresponding action takes the form

\begin{equation}
S=\frac{1}{k}\,\int {\rm d}^{4}x\,(A-B) . \label{013}
\end{equation}
Note that the Fierz tensor has dimensionality (length)$^{-1}$,
which is compatible with the fact that Einstein constant $k$ has
dimensionality (energy)$^{-1}$ (length)$^{-1}$. From now
on we set $k=1.$ Then,
\begin{equation}
\delta S=\int 2F^{\alpha \mu \nu }{}_{,\alpha }\,\delta
\varphi _{\mu \nu}\,d^{4}x . \label{018}
\end{equation}
Using the identity
\begin{equation}
F^{\alpha}{}_{\mu \nu,\alpha }=\frac{1}{2}\,F^{\alpha}{}_{(\mu\nu),\alpha}=
-\,\frac{1}{2}\,G^{(L)}{}_{\mu \nu },  \label{01888}
\end{equation}
we obtain
\begin{equation}
\delta S=-\int G^{(L)}{}_{\mu \nu }\,\delta\varphi^{\mu\nu}\,d^{4}x,
\label{018bis}
\end{equation}
where $G^{(L)}\mbox{}_{\mu \nu }$ is given in Eqn.(\ref{08}).
Thus, the action in Eqn.(\ref{013}) when written in the ER reads
\begin{equation}
S=-\int G^{(L)}{}_{\mu \nu }\,\varphi ^{\mu \nu }\,d^{4}x .
\label{0181bis}
\end{equation}

Let us consider now the massive case.
If we include a mass for the spin 2 field in the FR, the Lagrangian
takes the form
\begin{equation}
{\cal L}=A-B-\frac{m^{2}}{2}\,\left(\varphi _{\mu \nu }\, \varphi ^{\mu
\nu}-\varphi ^{2}\right),  \label{x055}
\end{equation}
and the EOM that follow are
\begin{equation}
F^{\alpha }{}_{(\mu \nu ),\alpha }
- m^{2}\,\left( \varphi_{\mu\nu}-\varphi\,\eta _{\mu \nu }\right) =0 ,
\label{mc1}
\end{equation}
or equivalently,
\[
G^{(L)}{}_{\mu \nu } +m^{2}\,\left( \varphi _{\mu \nu }-\varphi \,\eta _{\mu\nu }\right) =0.
\]
The trace of this equation gives
\begin{equation}
F^{\alpha }{}_{,\alpha }+ \frac{3}{2}\,m^{2}\,\varphi =0,
  \label{mc12}
\end{equation}
while the divergence of Eqn.(\ref{mc1}) yields
\begin{equation}
F_{\mu }=0.  \label{mc121}
\end{equation}
This result together with the trace equation gives $\varphi =0.$

In terms of the potential, Eqn.(\ref{mc121}) is equivalent to
\begin{equation}
\varphi _{,\,\mu }-\varphi ^{\epsilon }{}_{\mu \,,\epsilon }=0.
\label{mcd121}
\end{equation}
It follows that we must
have
\[
\varphi ^{\mu \nu }{}_{,\nu }=0.
\]
Thus we have shown that the original ten degrees of freedom (DOF)
of $F_{\alpha\beta\mu}$
have been reduced to five (which is the correct number
for a massive spin-2 field) by means of the  five constraints
\beq
\varphi ^{\mu \nu }{}_{,\nu }=0,\;\;\;\;\;\;\;\;\;\varphi = 0.
\label{fc}
\eeq

\section{Interaction of the spin-2 field with an electromagnetic field}
\label{int}

As discussed for instance in \cite{ds}, the minimal coupling
prescription
$\partial_\mu \rightarrow \partial_\mu - ieA_\mu$  is ambiguous in the case of a spin 2
field interacting with an EM field. The origin of this ambiguity is rooted, as in the case
of a curved background \cite{aragones71}, in the non-commutativity of the derivative
operator, which is manifest from
\beq
\varphi_{\alpha\beta ;\mu\nu} - \varphi_{\alpha\beta ;\nu\mu}
= ieA_{\nu\mu},
\label{comm}
\eeq
where $A_{\mu\nu}$ is the EM field,
and the semicolon is the covariant derivative $\partial_\mu - ieA_\mu$.
Let us review the argument in \cite{ds}, which starts from
the free Lagrangian for a charged spin 2 field in the ER:
$$
{\cal L} = \half \varphi^{*\mu\nu} G^{(L)}_{\mu\nu}
+ \half m^2(\varphi^{*\mu\nu}\varphi_{\mu\nu}- \varphi^* \varphi).
$$
The EOM that follow from this Lagrangian
are
\beqa
\Box (\varphi_{\mu\nu} - \eta_{\mu\nu} \varphi) + \varphi_{,\mu\nu} + \eta_{\mu\nu}
\varphi^{\alpha\beta}_{\;\;,\alpha\beta} -\varphi^{\alpha }_{(\nu ,\mu )\alpha} & \nonumber\\
+ m^2 (\varphi_{\mu\nu} - \eta_{\mu\nu} \varphi) & =0.
\eeqa
It is the term before the mass term
of this equation that leads to an ambiguity when minimal coupling
is adopted. In \cite{ds}, a one-parameter family of couplings was introduced, such that
$$
\varphi^{\alpha }_{(\nu ,\mu )\alpha} \rightarrow g \;\varphi^\alpha_{(\nu ;\mu ) \alpha}
+ (1-g)\varphi^\alpha_{(\mu ;\alpha\mu)}.
$$
By studying the constraints of the one-parameter theory, it was
shown in \cite{ds} that the only value of the gyromagnetic factor
$g$ that maintains the correct number of DOF is $g=1/2$. The
resulting EOM is
\beqa
\Box (\varphi_{\mu\nu} -\half
\eta_{\mu\nu} \varphi) + \varphi_{;(\mu\nu )} + \eta_{\mu\nu}
\varphi^{\alpha\beta}_{\;\; ;\alpha\beta} - \half \varphi^{\alpha }_{(\nu
;\mu )\alpha} &
\label{eomd} \\
-\half \varphi^{\alpha }_{(\nu ;\alpha\mu)}
+ m^2 (\varphi_{\mu\nu} - \eta_{\mu\nu} \varphi) & =0. \nonumber
\eeqa

Let us see how the minimal coupling procedure
affects the equations for the free field in the FR, given in Sect.\ref{fabc}.
First, in the presence of an EM field Eqn.(\ref{03}) transforms to
\beq
\stackrel{*}{F}{}^{\alpha (\beta\lambda )}_{\;\;\;\;\;\;\;\;;\alpha} = -\half ie
\stackrel{*}{A}{}^{\nu (\beta}\;\varphi^{\lambda )}_{\;\;\;\nu}.
\label{star}
\eeq
From this equation, the tensor $F_{\alpha\mu\nu}$ can be written as
\begin{equation}
2 F_{\alpha\mu\nu} = \varphi_{\nu [\alpha ;\mu ]}  +
F_{[\alpha } \,\eta_{\mu ]\nu},
\label{061}
\end{equation}
with
\begin{equation}
F_{\alpha} = \varphi_{;\alpha} - \varphi_{\alpha}{}^{\lambda}{}_{;\lambda}.
\label{051}
\end{equation}
If we start with the EOM for the charged spin-2 field in the absence of interactions in the
FR (Eqn.(\ref{mc1})), and
apply the minimal coupling procedure, we get
\beq
F^{\alpha }{}_{(\mu \nu );\alpha }
- m^{2}\,\left( \varphi_{\mu\nu}-\varphi\,\eta _{\mu \nu }\right) =0.
\label{eom1}
\eeq
There is no ambiguity then in the minimal substitution. In fact, using
Eqns.(\ref{061}) and (\ref{051}) in Eqn.(\ref{eom1}),
we get the equation derived in \cite{ds}
with $g=1/2$ ({\em i.e.} Eqn.(\ref{eomd})).
In other words,
the Fierz representation automatically gives a theory
with the correct number of degrees of freedom when the minimal coupling scheme is used.

Let us now give two constraints that follow from Eqn.(\ref{eom1}).
If we take the divergence on the index $\mu$ in Eqn.(\ref{eom1}), we get
\beq
 -\frac 3 2\; ie A^{\alpha\mu}F_{\alpha\mu\nu} + \half ie A^\mu_{\;\;\nu ,\alpha}
\varphi_\mu^{\;\alpha}  + m^2 F_\nu = 0,
\label{v1}
\eeq
for a  sourceless EM field $A_{\mu\nu}$.
Notice that in this constraint only first derivatives of $\varphi_{\mu\nu}$ appear (in
$F_\mu$).
Taking the divergence of
Eqn.(\ref{v1}) we obtain
\beq
ieA^{\alpha\mu}_{\;\;\;,\beta}\;F_{\alpha\mu}^{\;\;\;\beta}
- \frac 3 2 \left( m^4 - \frac 1 2\; e^2 A^2
 \right) \varphi + \frac 3 2 \;e^2 A^{\alpha}_{\;\mu}
A^{\mu\beta}\varphi_{\alpha\beta} = 0,
\label{v2a}
\eeq
where $A^2 = A_{\alpha\beta}A^{\alpha\beta}$.
Eqns.(\ref{v1}) and (\ref{v2a}) correspond to the
free-case equations (\ref{fc}). They reduce the number of DOF to five,
and are necessary for the compatibility of the system.
Note that a
remarkable cancellation has happened: no second derivatives of $\varphi_{\mu\nu}$
are present in this second constraint.
It is precisely the absence
in the constraints of second derivatives w.r.t time that guarantees that only physical
degrees of freedom propagate. Armed with the EOM (\ref{eom1}),
we shall study in the next section the causal properties of massive spin 2
particles in an EM background.

\section{Causality in spin 2 fields interacting with an EM background}
\label{caus}

In this section it will be shown, using the FR,
that the propagation of a massive spin 2 field
in an EM background is causal. We shall recourse to
the well-known method of the characteristics,
which is in fact equivalent to the infinite-momentum limit of the
eikonal approximation \cite{hada}. To set the stage for the calculation, let us
put together the equations we shall use. They are the EOM,
\beq
F^{\alpha }{}_{(\mu \nu );\alpha }
 -m^{2}\,\left( \varphi_{\mu\nu}-\varphi\,\eta _{\mu \nu }\right) =0,
\label{eom11}
\eeq
its trace,
\begin{equation}
F^{\alpha }{}_{;\alpha }-\frac{3}{2}\,m^{2}\,\varphi =0,
\label{traco}
\end{equation}
and the two constraints
\beq
-\frac 3 2 \;ie A^{\alpha\mu}F_{\alpha\mu\nu} + \half ie A^\mu_{\;\;\nu ,\alpha}
\varphi_\mu^{\;\alpha}    + m^2 F_\nu = 0,
\label{v11}
\eeq
and
\beq
ieA^{\alpha\mu}_{\;\;\;,\beta}\;F_{\alpha\mu}^{\;\;\;\beta}
- \frac 3 2 \left( m^4 - \frac 1 2\; e^2 A^2 \right) + \frac 3 2 \;e^2 A^{\alpha}_{\;\mu}
A^{\mu\beta}\varphi_{\alpha\beta} = 0.
\label{v2}
\eeq
To these, we must add some properties of the Fierz tensor:
\beq
F_{\alpha\mu\nu}+ F_{\mu\alpha\nu} =0,
\label{antys}
\eeq
\begin{equation}
F_{\alpha\mu\nu} + F_{\mu\nu\alpha} + F_{\nu\alpha\mu} = 0,
\label{cyclic}
\end{equation}
and
\beq
\stackrel{*}{F}{}^{\alpha (\beta\lambda )}_{\;\;\;\;\;\;\;\;;\alpha} = -\half ie
\stackrel{*}{A}{}^{\nu (\beta}\;\varphi^{\lambda )}_{\;\;\;\nu}.
\label{star2}
\eeq

Let $\Sigma$ be the surface of discontinuity defined by the equation
$$\Sigma(x^{\mu}) = {\rm constant}.$$
The discontinuity of a function $J$ through $\Sigma$ will be represented by
$[J]_\Sigma$, and its definition is
$$
[J]_\Sigma \equiv \lim_{\delta\rightarrow 0^+} \left( \left. J\right|_{\Sigma +\delta}
- \left. J \right|_{\Sigma - \delta}\right) .
$$
We shall assume that $F_{\alpha\mu\nu}$ is continuous through the surface $\Sigma$
but its first derivative is not:
\beq
[F_{\alpha\mu\nu}]_\Sigma = 0, \;\;\;\;\;\;\;\;[F_{\alpha\mu\nu ;\lambda}]_\Sigma =
f_{\alpha\mu\nu}k_\lambda.
\label{disc}
\eeq
From the discontinuity of the EOM (\ref{eom11}) we learn that
\beq
f^{\mu (\alpha \beta)}k_\mu =0.
\label{d1}
\eeq
Taking the derivative of Eqn.(\ref{cyclic}) results in
\beq
k^\alpha f_{\alpha\mu\nu} + k^\alpha f_{\nu\alpha\mu} + k^\alpha
f_{\mu\nu\alpha} =0.
\label{d3}
\eeq
This equation, together with Eqns.(\ref{disc}) and (\ref{d1}) tells us that
the contraction of $k$ with $f$ is zero on any index of $f$.
The trace equation (\ref{traco}) gives
\beq
f^\mu k_\mu =0.
\label{d4}
\eeq
Eqn.(\ref{star}) can be written as
\begin{eqnarray}
{{F_{\alpha\beta}}^{\lambda}}{}_{,\mu} + {{F_{\beta\mu}}^{\lambda}}%
{}_{,\alpha} + {{F_{\mu\alpha}}^{\lambda}}{}_{,\beta} -\frac{1}{2}\;
\delta^{\lambda}_{\;\alpha} F_{[\mu ,\beta ]} + &  \\
 - \frac{1}{2} \;\delta^{\lambda}_{\;\mu} F_{[\beta ,\alpha ]}
  - \frac{1}{2} \;\delta^{\lambda}_{\;\beta} F_{[\alpha ,\mu ]}  = & -\half ie
\stackrel{*}{A}{}^{\nu (\rho}\;\varphi^{\lambda )}_{\;\;\;\nu}.\nonumber
\label{star22}
\end{eqnarray}
Notice that the r.h.s. is continuous. Taking the discontinuity of this equation,
multiplying by $k^\mu$ and $f_\lambda$, and using Eqn.(\ref{d1}), we get that
\beq
f_{\alpha\beta\lambda}f^\lambda k^2 = 0.
\label{d5}
\eeq
We shall assume for the time being that $k^2\neq 0$.
The discontinuity of the derivatives of the constraints Eqns.({\ref{v11})
and (\ref{v2}) give
\beq
A_{\alpha\beta ,\mu}f^{\alpha\beta\mu} = 0,
\label{dv1}
\eeq
\beq
\frac 3 2 \;ie\; A^{\alpha\beta}f_{\alpha\beta\mu}-m^2f_\mu = 0.
\label{dv2}
\eeq
From Eqns.(\ref{d5}) and (\ref{dv2}) we deduce that
\beq
f_\mu f^\mu = 0.
\label{d6}
\eeq
Note that {\em all} the equations that resulted from taking the discontinuity
({\em i.e.} Eqns.(\ref{d1})-(\ref{d4}) and (\ref{d5})-(\ref{dv2}))
depend only
on $f_{\mu\nu\alpha}$ and its trace. Taking the discontinuity of
the derivative of $F_{\mu\nu\alpha}$ we get that
$$
[F_{\alpha\mu\nu ,\lambda}]_\Sigma = f_{\alpha\mu\nu}k_\lambda
$$
where
\beq
2f_{\alpha\mu\nu} = \epsilon_{\nu\alpha}k_{\mu}- \epsilon_{\nu\mu}k_{\alpha}
+ f_\alpha \eta_{\mu\nu} - f_\mu \eta_{\alpha\nu},
\label{discf}
\eeq
and
$$
f_\alpha = \epsilon k_\alpha - \epsilon _\alpha^\beta k_\beta.
$$
Consequently the equations that follow from taking
the discontinuity are invariant under the
transformation
\beq
\epsilon'_{\mu\nu} = \epsilon_{\mu\nu} + \Lambda k_\mu k_\nu,
\label{trafo}
\eeq
where $\Lambda $ is an arbitrary
function of the coordinates \footnote{Notice that this, as a direct calculation shows,
is {\em not}
a symmetry of the massless theory in the presence of
a nonzero background.}. This equation
implies that
\beq
\epsilon' = \epsilon + \Lambda k^2.
\label{moda}
\eeq
Now, this symmetry implies that observable quantities depend on the
gauge choice {\em unless $k^2 =0$}.
Let us take for instance $X_\mu = \epsilon_{\mu\nu}k^\nu$.
From Eqn.(\ref{d4}),
\beq
X'_\mu k^\mu = X_\mu k^\mu + (k^2)^2.
\label{kx}
\eeq
It follows that the projection of the polarization
in the direction of $k_\mu$
is not a gauge-invariant quantity.
Another quantity
that is not gauge invariant is the norm $X_\mu X^\mu$, which transforms as
\beq
X'^2= X^2 + ( 2  \epsilon + \Lambda) \Lambda (k^2)^2.
\eeq
We see that a spacelike (actually, a non-null) $k_\mu$ entails an unacceptable
dependence of observable quantities with the gauge choice. This dependence
disappears only when $k^2 = 0$.
Summing up,
the propagation of spin two fields in an EM background
is causal, with the characteristics governed by
the equation $k^2 = 0$.

\section{Conclusion}

We have given a summary of the Fierz representation for a spin 2 field, both in the
free case and for the interaction with an electromagnetic background. This representation
has some advantages over the Einstein representation. In particular,
it was shown that, while the ER of a spin-2 field in an electromagnetic background
has an inherent ambiguity related to the order of the derivatives when the
minimal coupling procedure is applied, the Fierz representation
is free from this difficulty.
Another advantage
of this representation is that it is similar to that used in Electromagnetism, and
then we can profit from work already done in this area for instance in construction
nonlinear theories for the spin 2 field
\cite{mio}. More importantly, the use of the Fierz representation
has paved the way to a clean proof of the causality in the propagation of spin 2 fields in
the presence of an electromagnetic field, thus showing that previous claims about
noncausal propagation were mistaken.

To close, we would like to point out that
in the issue of causality, the use of the Fierz representation is not mandatory. A
closer look to the relevant equations in the Einstein representation (for instance Eqns.
(64) and (66) in \cite{ds}, without choosing a timelike $k_\mu$) shows that the gauge
invariance given by Eqn.(\ref{trafo}) is present there too. However, it is important
to remark that the gauge invariance of the equations
for the discontinuity (which went unnoticed before) is clearly displayed
in the Fierz representation.

\section*{Appendix 1}


We shall be concerned here
with the gauge invariance of
Eqn.(\ref{014bis}) under the map
\begin{equation}
\varphi_{\mu\nu} \rightarrow \tilde{\varphi}_{\mu\nu} = \varphi_{\mu\nu} +
\Lambda_{\mu ,\nu} + \Lambda_{\nu , \mu} . \label{30011}
\end{equation}
Although the field $F_{\alpha\beta\mu}$ is invariant under this map only
if the vector $\Lambda_{\mu}$ is a
gradient, it is important to realize that
the dynamics is invariant even when $\Lambda$ is not a gradient. Indeed, we
have
\begin{equation}
\delta F_{\alpha\beta\mu} \equiv \tilde{F}_{\alpha\beta\mu} -
F_{\alpha\beta\mu} = \frac{1}{2} \, X_{\alpha\beta\mu}{}^{\,\lambda}{}_{\,
,\lambda},
\end{equation}
where
\begin{eqnarray}
X_{\alpha\beta\mu}{}^{\lambda} &\equiv&(\Lambda_{\alpha ,\beta} -
\Lambda_{\beta ,\alpha}) \delta^{\lambda}_{\mu} +
[ \Lambda^{\sigma}{}_{,\sigma} \delta^{\lambda}_{\alpha} -
\Lambda_{\alpha}{}^{,\lambda}] \eta_{\beta\mu} \nonumber \\
&-&[\Lambda^{\sigma}{}_{,\sigma} \delta^{\lambda}_{\beta} -
\Lambda_{\beta}{}^{,\lambda} ] \eta_{\alpha\mu}. \label{119}
\end{eqnarray}
Then it follows that
\begin{equation}
2\delta F_{\alpha} = X_{\alpha}{}^{\, \lambda}{}_{\, ,\lambda},
\end{equation}
with
\[
X_{\alpha}{}^{\,\lambda} \equiv X_{\alpha\beta}{}^{\,\beta\lambda}.
\]

As a consequence of this transformation, the invariants $A$ and
$B$ change in the following way:
$$
\delta A =
F^{\alpha\beta\mu}X_{\alpha\beta\mu}{}^{\,\lambda}{}_{\,
,\lambda},\;\;\;\;\;\;\;\; \delta B =
F^{\alpha}X_{\alpha}{}^{\,\lambda}{}_{,\lambda}.
$$
Note that $X_{\alpha\beta\mu}{}^{\lambda}$ is {\bf not} cyclic
in the indices $(\alpha\beta\mu )$, but the quantity
$X_{\alpha\beta\mu}{}^{\,\lambda}{}_{,\lambda}$
has such cyclic property:
\begin{equation}
X_{\alpha\beta\mu}{}^{\,\lambda}{}_{,\lambda} +
X_{\beta\mu\alpha}{}^{\,\lambda}{}_{,\lambda} +
X_{\mu\alpha\beta}{}^{\lambda}{}_{,\lambda} =0.
\end{equation}
It is straightforward to show the associated identities:
\begin{equation}
X^{\alpha\beta\mu\lambda}{}_{,\lambda \alpha} = 0
\label{id}
\end{equation}
\begin{equation}
X^{\alpha\beta\mu\lambda}{}_{,\lambda \mu} = 0
\end{equation}
\begin{equation}
{X^{\alpha\lambda}{}_{,\alpha\lambda} = 0}.
\end{equation}
Thus,
\begin{equation}
\delta A = [\varphi^{\mu\alpha ,\beta} + F^{\alpha}\,\eta^{\mu\beta}] \
X_{\alpha\beta\mu}{}^{\lambda}{}_{\,,\lambda}  \nonumber,
\end{equation}
or, equivalently,
\begin{equation}
\delta A = \varphi^{\mu\alpha ,\beta} X_{\alpha\beta\mu}{}^{\,\lambda}{}_{\,
,\lambda} + F^{\alpha} X_{\alpha}{}^{\,\lambda}{}_{\, ,\lambda}.
\end{equation}
Then,
\begin{equation}
\delta (A - B) = \varphi_{\mu\alpha ,\beta} X^{\alpha\beta\mu\lambda}{}_{\,
,\lambda},
\end{equation}
and
\begin{equation}
\int \varphi_{\mu\alpha ,\beta} X^{\alpha\beta\mu\lambda}{}_{\, ,\lambda} = \int
div - \int \varphi_{\mu\alpha} X^{\alpha\beta\mu\lambda}{}_{ ,\lambda \beta},
\end{equation}
so that, because of (\ref{id}),
\begin{equation}
\int \delta (A - B) = 0.
\end{equation}

This shows that the transformation
\[
F_{\alpha\beta\mu}\rightarrow F_{\alpha\beta\mu} +
X_{\alpha\beta\mu}{}^{\,\lambda}{}_{\,,\lambda},
\]
for $X_{\alpha\beta\mu}{}^{\lambda}$ given in equation (\ref{119}), leaves
the dynamics invariant.

\section{Acknowledgements}

This work was partially supported by the Brazilian agency Conselho Nacional
de Desenvolvimento Cient\'{\i}fico e Tecnol\'ogico (CNPq), and by FAPERJ.

\end{document}